# Multi-branch Cascaded Swin Transformers with Attention to k-space Sampling Pattern for Accelerated MRI Reconstruction

Mevan Ekanayake[a,b,1], Kamlesh Pawar[a], Mehrtash Harandi[b], Gary Egan[a,c], and Zhaolin Chen[a,d]

[a]*Monash Biomedical Imaging, Monash University, Clayton, VIC 3800 Australia*
[b]*Department of Electrical and Computer Systems Engineering, Monash University, Clayton, VIC 3800 Australia*
[c]*School of Psychological Sciences, Monash University, Clayton, VIC 3800 Australia*
[d]*Department of Data Science and AI, Monash University, Clayton, VIC 3800 Australia*

## Abstract

Global correlations are widely seen in human anatomical structures due to similarity across tissues and bones. These correlations are reflected in magnetic resonance imaging (MRI) scans as a result of close-range proton density and T1/T2 parameters in tissues. Furthermore, to achieve accelerated MRI, k-space data are undersampled which causes global aliasing artifacts. Convolutional neural network (CNN) models are widely utilized for accelerated MRI reconstruction, but those models are limited in capturing global correlations due to the intrinsic locality of the convolution operation. The recent self-attention-based transformer models are capable of capturing global correlations among image features, however, the current contributions of transformer models for MRI reconstruction are minute. The existing contributions mostly provide CNN-transformer hybrid solutions and rarely leverage the physics of MRI. In this paper, we propose a physics-based stand-alone (convolution free) transformer model titled, the Multi-branch Cascaded Swin Transformers (McSTRA) for accelerated MRI reconstruction. McSTRA combines several interconnected MRI physics-related concepts with the transformer networks: it exploits global MR features via the shifted window self-attention mechanism; it extracts MR features belonging to different spectral components separately using a multi-branch setup; it iterates between intermediate de-aliasing and k-space correction via a cascaded network with data consistency in k-space and intermediate loss computations; furthermore, we propose a novel positional embedding generation mechanism to guide self-attention utilizing the point spread function corresponding to the undersampling mask. With the combination of all these components, we propose a consolidated robust solution for MRI reconstruction which shows improved performance by significantly outperforming state-of-the-art MRI reconstruction solutions both visually and quantitatively. Further experiments demonstrate the superior capabilities of McSTRA in combating adversarial conditions such as higher accelerations, noisy data, different undersampling protocols, out-of-distribution data, and abnormalities in anatomy.

[1]Corresponding author.
E-mail address: mevan.ekanayake@monash.edu

# 1 Introduction

The outstanding soft tissue contrast and flexibility make magnetic resonance imaging (MRI) a comprehensive diagnostic tool for a broad range of disorders, including musculoskeletal, neurological, and oncological diseases. Compared with other modalities, MRI is unique in producing non-invasive measures for imaging anatomy, diffusion process, and functions in the human body (Nishimura, 1996). However, the long data acquisition time is a significant issue in MRI, which leads to patient discomfort, image artifacts, and high examination costs (Zbontar et al., 2019). Image reconstruction in MRI is an inverse process that reconstructs images with diagnostic quality from acquired MR data (or k-space). A trade-off between image quality and acquisition time is often made by undersampling k-space to accelerate the data acquisition process, which is commonly known as accelerated MRI reconstruction.

Undersampling k-space violates the Nyquist-Shannon sampling theorem thereby introducing aliasing artifacts. These artifacts must be eliminated during the course of the image reconstruction process. Early solutions involved linear analytic methods such as partial Fourier encoding, sensitivity encoding (SENSE) (Pruessmann et al., 1999), simultaneous acquisition of spatial harmonics (SMASH) (Sodickson and Manning, 1997), and generalized auto calibrating partially parallel acquisitions (GRAPPA) (Griswold et al., 2002), which incorporates prior knowledge of the imaging systems as well as k-space correlations. Later, the reconstruction methodologies drifted toward solving optimization problems using non-linear iterative algorithms which incorporated the physics of the imaging system and as well as regularization of priors such as sparsity (Chaâri et al., 2011), low rank (Haldar and Zhuo, 2016), manifold (Wachinger et al., 2012), total variation (Block et al., 2007; Chen et al., 2018, 2017), and dictionary learning (Weller, 2016). However, these methods are limited in both model capacity and model parameter tuning, hence offering inadequate flexibility resulting in suboptimal performance.

Driven by the explosion in artificial intelligence, deep learning (DL) methods have shown promising results for accelerated MRI reconstruction (Hammernik et al., 2018; Wang et al., 2016; Zhu et al., 2018) due to their flexibility and capacity. These methods are capable of learning complex patterns from a large amount of training data. Many DL-based MRI reconstruction methods are subsequently developed using convolutional neural network (CNN) models which can be studied under two main categories (Wang et al., 2021), i.e. data-driven end-to-end DL methods which map the low-quality undersampled images to high-quality references (Hyun et al., 2018; Pawar et al., 2021, 2019; Wang et al., 2016; Yang et al., 2018) and physics-constrained DL methods which iteratively solve for an inverse problem (Aggarwal et al., 2019; Schlemper et al., 2018; yang et al., 2016).

Global correlations are often observed in human anatomical structures as a result of the similarity across different tissues and bones. These are reflected in MRI scans as a result of close-range proton density and T1/T2 parameters. In a random under-sampling setting, the point spread function (PSF) corresponding to the under-sampling mask involves more voxels that are aliased together to produce incoherent aliasing artifacts. These are observed globally within the field of view in image space (See Section 3.5). As a result, exploiting global correlations could be highly beneficial in accomplishing improved image quality in MRI such as in the denoising application using non-local means filtering (Manjón et al., 2008) and BM3D filtering (Dabov et al., 2006). The CNN models are suboptimal in capturing these global image correlations due to the intrinsic locality of the convolution operation caused by the narrow receptive fields of the convolution kernels. This issue could be mitigated by stacking more convolution layers sequentially but that would increase the computational burden and the amount of training required.

The transformer models have recently shown significant improvement in capturing global correlations via the inherent self-attention mechanism in many natural language processing (NLP) and computer vision (CV) tasks. With respect to CV-based applications, the image is simply divided into patches; the patches are converted into vectorized tokens; and the tokens are subjected to self-attention. Even though transformer models are generally computationally expensive compared to the conventional CNN models, this patch-based strategy in transformers helps to reduce the computational burden by reducing a single patch to a token rather than considering each pixel individually. Self-attention aggregates global correlations of the entire image at once, rather than confining to a local kernel. Despite the rapid success, the current applications of transformer models for accelerated MRI reconstruction are minute and suboptimal. There have been several attempts, however, these models utilize convolution layers at one stage or another within the model architecture in order to leverage local features which limits global feature extraction. Korkmaz and colleagues proposed two unsupervised learning-based models centered around deep image prior and adversarial learning, respectively. Korkmaz relied on convolution modules to further process the attention-based contextual representations while Korkmaz et al. (2022) relied on convolutional layers for upsampling features. Huang et al. (2022) proposed a Swin transformer-based model, nevertheless, it utilizes convolutional layers to map from the input image space to higher dimensional feature space. Feng et al. (2021) and Yan et al. (2021) proposed transformer-based models for joint MRI reconstruction and super-resolution, yet these models made use of convolutional layers at early stages for shallow feature extraction.

In this paper, we aim to develop a novel physics-based transformer model referred to as McSTRA (stands for Multi-branch Cascaded Swin Transformers) for accelerated MRI reconstruction by combining several interconnected physics-based strategies. Our model is free of convolution layers and to the best of our knowledge, the first study to utilize the Swin-Unet (Cao et al., 2021) incorporated physics knowledge for accelerated MRI reconstruction. The Swin-Unet model hierarchically extracts features at different spatial resolutions while incorporating the shifted window self-attention mechanism allowing accurate reconstruction of global features in all resolution levels. First, McSTRA extracts MR features belonging to different spectral components by decomposing the k-space into low-frequency features (corresponding to anatomical and structural information) and high-frequency features (corresponding to edge features and resolution information) and reconstructing them separately using a multi-branch setup. Then, it iterates between intermediate de-aliasing and k-space correction via a cascaded network with data consistency (DC) in k-space and intermediate loss computations in order to incrementally reconstruct images. Finally, it performs an image-to-image mapping in order to reconstruct the final magnitude MR image. We also propose a novel positional embedding generation mechanism to guide self-attention layers within the transformer models. This mechanism exploits the relationship between the PSF of the sampling protocol and the pattern of the aliasing artifacts within the field of view to generate an instance-dependent positional embedding. Overall, we summarize below the key novelties and contributions of our work.

- We propose a novel positional embedding generation scheme based on the point-spread function of the sampling mask to guide the self-attention mechanism in transformers for robust MRI Reconstruction.
- We incorporate a multi-branch k-space partitioning mechanism in the DL setting which extracts low- and high-frequency components separately in order to recover fine features of MR images.
- Our proposed framework is convolution free and utilizes only self-attention layers, hence our work validates the potential of transformer-based frameworks in the domain of MRI reconstruction supported by a comprehensive performance comparison carried out with state-of-the-art DL reconstruction methods.

The contribution of the overall framework proposed in this paper is a combination of the interconnected MRI physics-related concepts mentioned above, as well as the contributions in individual components including the multi-branch feature extraction, cascaded network and intermediate loss computations, reconstruction tail, and our approach of sampling mask-guided positional embeddings. We present the consolidated solution, "McSTRA" for accelerated MRI reconstruction which qualitatively as well as quantitatively demonstrates superior performance over the previous state-of-the-art solutions. The results demonstrate comprehensive performance improvements including in the presence of adversarial conditions such as higher accelerations, noisy data, different undersampling protocols, out-of-distribution data, and in the presence of anatomical abnormalities. We believe this work will contribute to reinforcing the establishment of transformer-based models for high-quality MRI reconstruction.

The remainder of the paper is arranged as follows. Section 2 describes the literature related to the underlying concepts of McSTRA. Section 3 presents several theoretical formulations to support the functionality of the constituent components of McSTRA. In Section 4, the novel McSTRA model is introduced. Section 5 shows empirical results, including quantitative and qualitative evaluations on the large-scale fastMRI dataset followed by ablation studies to comprehend the behavior of McSTRA. Section 6 summarizes the paper and describes the limitations of our work along with the future directions.

## 2 Related Work

This Section summarizes the previous works that align with the underlying concepts of McSTRA.

**Cascaded Deep Learning Models for MRI Reconstruction**. Several previous works proposed DL models for MRI reconstruction that consist of cascaded CNN blocks and DC blocks that perform k-space correction (Schlemper et al., 2018; Souza et al., 2019; Zheng et al., 2019). In Schlemper et al. (2017, 2018), a cascaded DL model consisting of convolution layers was proposed for accelerated MRI reconstruction in cardiac MR. The cascaded network consists of CNN models interleaved DC layers and data sharing layers where the data sharing layers are utilized for dynamic sequences. The underlying principle of the cascaded network is to iterate between the intermediate de-aliasing and k-space correction. The 2D slice reconstruction in this study showed considerable improvement in reconstruction over the traditional compressed sensing methods. Zheng et al. (2019) proposed a Cascaded Dilated Dense Network (CDDN) for MRI reconstruction which contains multiple sub-network iterations and after each sub-network, a two-step data consistency (TDC) is performed on the k-space. The experiments were conducted on the cardiac dataset and the fastMRI dataset where significant improvements were observed. Apart from that, Kocanaogullari and Eksioglu (2019), and Souza et al. (2019) also adopted the cascaded structure in DL setting for MRI reconstruction.

**Transformers**. Self-attention-based transformer models which exploit the global correlations in data have shown massive improvement in NLP tasks such as machine translation, text classification, and visual question answering (Galassi et al., 2021; Vaswani et al., 2017). Following this success, the CV community has shown considerable interest in adopting transformer models for various vision tasks over the last couple of years. For example, Google introduced the Vision Transformer which was able to outperform all the state-of-the-art CNN-based models in image classification (Dosovitskiy et al., 2021).

The recent Swin transformer (Liu et al., 2021) was able to outperform many state-of-the-art methods in image segmentation and object detection. The Swin transformer is a transformer variant which not only operates on image patches but also models global correlations via a shifted window mechanism (See Fig. 1(c)). This type of partitioning mechanism will enable the model to capture global correlations for an effective reconstruction. Several transformer models such as Swin-Unet (Cao et al., 2021), TransUNet (Chen et al., 2021), and Medical Transformer (Valanarasu et al., 2021) have been proposed specifically to solve medical image analysis tasks such as segmentation and detection (Peiris et al., 2022). Swin-Unet is an Unet-like pure transformer network proposed for medical image segmentation. Similar to the conventional U-Net architecture, Swin-Unet consists of an encoder, a bottleneck, and a decoder containing several Swin transformer blocks as can be seen in Fig. 3(c).

**Transformer-based MRI Reconstruction**. Recently, there have been several transformer-based DL models proposed for MRI reconstruction. The SwinMR model (Huang et al., 2022) couples parallel imaging with the Swin transformer for multi-channel MRI reconstruction. Their experiments on Calgary Campinas multi-channel brain dataset and the Multi-modal Brain Tumor Segmentation Challenge 2017 (BraTS17) dataset demonstrated superior performance over CNN-based and GAN-based MRI reconstruction methods. SwinMR also showed excellent robustness to various undersampling trajectories and noise levels.

Korkmaz et al. (2022, 2021) proposed unsupervised leaning-based transformer models for MRI reconstruction. Their work centered around deep image prior (DIP) and adversarial learning. The DIP-based model (Korkmaz et al., 2021) referred to as GVTrans progressively maps low-dimensional noise and latent variables onto MR images via cascaded blocks of cross-attention ViTs. The adversarial learning-based method (Korkmaz et al., 2022) referred to as SLATER, embodies a deep adversarial network with cross-attention transformers to map noise and latent variables onto coil-combined MR images. Their experiment on the IXI brain dataset and the fastMRI brain dataset showed improved image quality compared to CNN-based reconstructions.

Several works proposed transformer-based models for joint MRI reconstruction and super-resolution. Feng et al. (2021) focused on multi-task learning which involves sharing structural information between higher-quality and super-resolved reconstructions and their experiments on the IXI brain MRI dataset showed reduced blurring and artifacts. Yan et al. (2021) proposed a Swin transformer-based model referred to as SMIR which focuses on two simultaneous tasks: multi-level feature extraction and image reconstruction. SMIR incorporates both spatial and frequency domain losses and their experiments display improved performance under a wide range of acceleration factors on the HCP brain dataset.

Feng et al. (2022) proposed a unified transformer framework referred to as MTrans for multi-modal MR imaging. MTrans utilizes a cross-attention module, which extracts and transfers features from an auxiliary modality to a target modality. Their experiments on fastMRI knee and brain MRI datasets validate improved reconstruction performance.

**Positional Embeddings**. Transformer networks are fundamentally built on the concept of multi-head self-attention. This self-attention operation in transformers unlike CNNs involves fully connected layers, hence permutation-invariant, i.e., it does not have awareness of the order of the input token sequence. This could harm the learning process of transformers, especially in the image domain since the order of the sequence is important for capturing its content. In order to incorporate the order of sequence, initial works proposed fixed positional embeddings (Vaswani et al., 2017) with sinusoidal functions and learnable positional embeddings (Dosovitskiy et al., 2021) that were added to input features before subjecting to the self-attention operation.

Several learnable positional embedding strategies were proposed in Dosovitskiy et al. (2021) such as 1-dimensional positional embedding, 2-dimensional positional embedding, and relative positional embedding (also discussed in Shaw et al. (2018)). Influenced by the success of incorporating positional embeddings, several other variants were proposed later. In Liu et al. (2020), an efficient position encoding scheme called FLOATER was introduced where position information was modeled with a continuous dynamical model. In Chu et al. (2021), a conditional positional embedding strategy was introduced where a position encoding generator produces positional embeddings conditioned on the input.

**Decomposition of k-space**. The k-space signals are well known to be non-uniformly distributed in terms of spectral energy. The magnitude tends to be larger in low-frequency bands (or center region) and contains more anatomical and structural information. The magnitude is small in high-frequency bands (or the outer region) and contains edge features and resolution information (Nguyen-Duc et al., 2019; Wu et al., 2016). Nevertheless, the high-frequency bands contain diagnostically important information (He et al., 2019). Many DL-based models enforce a uniform loss function over the entire field of view and k-space which results in imbalance training for the high frequencies in k-space due to their intrinsic low magnitude. Such imbalance can cause further loss of resolution.

Improved characterization of k-space has demonstrated improved accuracy in MRI reconstruction, e.g., compressed sensing MRI (Nguyen-Duc et al., 2019; Sung and Hargreaves, 2013; Yang et al., 2015), spread spectrum MRI (Pawar et al., 2020, 2015), self-supervision (Yaman et al., 2020). Further, The study by Sun et al. (2019) explicitly demonstrated the significance of sustaining high-frequency k-space information where the authors propose a subspace decomposition approach which is accomplished by using complementary filter banks referred to as HoriVert. The images are reconstructed in complementary subspaces belonging to high- and low-frequency decomposition in k-space individually using off-the-shelf algorithms, and finally fused together to obtain the final reconstruction. Similar approaches have been adopted in several other works (Faris, 2021; Faris et al., 2021; He et al., 2019).

## 3  Theory

This Section provides several theoretical formulations which we utilize to support the functionality of the constituent components of McSTRA. Throughout the remainder of the paper, we utilize the variable $x$ and its variants to denote MR data in image space, the variable $y$ and its variants to denote MR data in k-space, and the variable $z$ and its variants to denote intermediate latent MR features.

### 3.1  MRI Reconstruction

Let $x \in \mathbb{C}^N$ be the complex-valued vector representation of an MR image and $y \in \mathbb{C}^N$ be the fully sampled measurements in k-space. Then, the forward model of MRI acquisition can be presented as $y = Fx$, where $F \in \mathbb{C}^{N \times N}$ denotes the Fourier encoding matrix. Thus, in a fully sampled scenario, a clean MR image as seen in Fig. 1(a) can be reconstructed by $x = F^H y$, where $F^H$ is the inverse Fourier encoding matrix. Note that the above represents single-channel (or channel combined) data acquisition whereas, in multi-channel data acquisition, a coil sensitivity matrix should be included. Accelerated MRI can be achieved by sampling only a subset of k-space samples, $\hat{y}$. This is usually modelled using a binary sampling mask, $M \in \{0,1\}^N$ i.e., $\hat{y} = M \odot y$, where $M$ takes a value of 1 at locations where the k-space is sampled and 0 elsewhere, and $\odot$ represents the Hadamard product. A directly reconstructed image, $\hat{x} = F^H \hat{y}$ will have global aliasing artifacts as seen in Fig. 1(b).

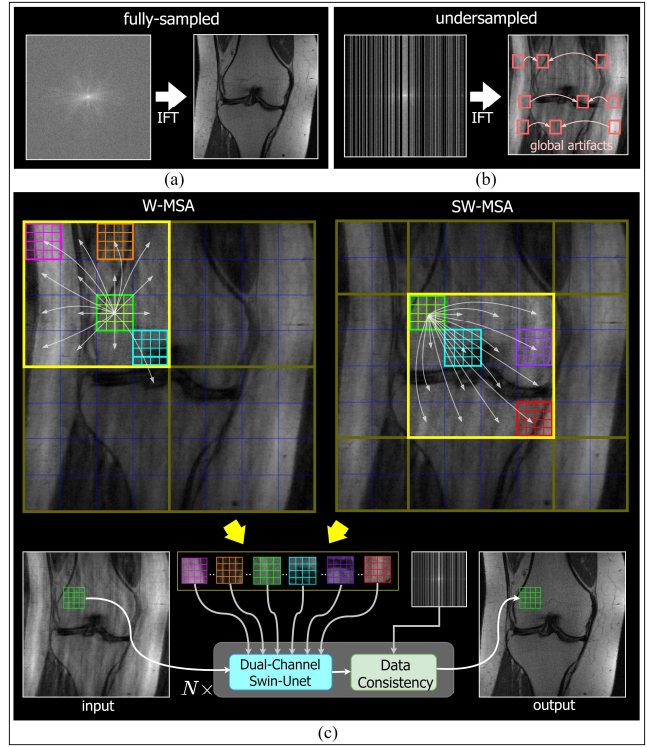

**Fig. 1.** (a) MRI reconstruction from fully sampled k-space data. (b) MRI reconstruction from undersampled k-space data causing global artifacts. (c) Window multi-head self-attention (W-MSA) and shifted window multi-head self-attention (SW-MSA) which are administered by two consecutive transformers within the Swin transformer used in McSTRA. The figure depicts how an image patch (light green) is able to broaden its receptive field and capture global features via the shifted window mechanism and self-attention.

### 3.2 Deep Learning Formulation

In order to reconstruct $x$ from $\hat{y}$, the following regularized optimization can be used (Schlemper et al., 2018):

$$x^* = \underset{x}{\mathrm{argmin}}\{\mathcal{R}(x) + \lambda\|\hat{y} - M \odot (Fx)\|_2^2\} \qquad (1)$$

where the $\mathcal{R}(\cdot)$ performs regularization, e.g. based on sparsity priors (Lustig et al., 2007; Ravishankar and Bresler, 2011) or network priors (Wang et al., 2016). The second term ensures data fidelity and $\lambda \in \mathbb{R}$ allows the balance of data fidelity based on the acquisition noise level (Schlemper et al., 2018). Most of the DL-based reconstruction methods are data-driven and end-to-end trainable, therefore, learn a complex high-dimensional regularization function through training data itself. The formulation of the DL-based reconstruction can be presented as follows:

$$x^* = \underset{x}{\mathrm{argmin}}\{\|x - f_{\mathrm{DL}}(\hat{x}|\theta)\|_2^2 + \lambda\|\hat{y} - M \odot (Fx)\|_2^2\} \qquad (2)$$

where $f_{DL}$ is the forward DL model parameterized by θ which takes an image with artifacts, $\hat{x}$, as input and produces a reconstruction as output. Given a large set of training data, the DL model can be trained using a predefined loss function such as $\ell_1$-norm (Zbontar et al., 2019), $\ell_2$-norm (Schlemper et al., 2018), categorical cross-entropy (Pawar et al., 2019), or structural similarity (Zbontar et al., 2019) -based loss functions.

A closed-form solution to the optimization problem in Eq. (2) can be presented as follows (Wang et al., 2016):

$$y_{rec}(j) = \begin{cases} y_c(j) & \text{if } j \notin \Omega \\ \dfrac{y_c(j) + \lambda \hat{y}(j)}{1+\lambda} & \text{if } j \in \Omega \end{cases} \quad (3)$$

where $y_c = Fx_c = Ff_{DL}(\hat{x}|\theta^*)$ is the k-space of the reconstructed image, $x_c$. Parameters, $\theta^*$ are the optimized model parameters. $\Omega$ indicates the subset of indices sampled from the k-space. The final reconstructed image is given as $x_{rec} = F^H y_{rec}$.

## 3.3 Cascaded Network

As discussed in Section 2, cascaded networks for MRI reconstruction consist of cascaded DL blocks and DC blocks that perform the k-space correction. The forward pass of the DC block can be expressed in matrix form as follows:

$$f_{DC}(x_c, y; \lambda) = F^H \Lambda F x_c + \frac{\lambda}{1+\lambda} F^H (M \odot y) \quad (4)$$

Here, $\Lambda$ is a diagonal matrix where the diagonal element at location $(k, k)$ is given by $\Lambda_{(k,k)} = \begin{cases} 1 & \text{if } k \notin \Omega \\ \frac{1}{1+\lambda} & \text{if } k \in \Omega \end{cases}$ where $k$ is the index. $\Omega$ indicates which k-space measurements have been sampled in $y$. The underlying principle of this cascaded model is that the output of one CNN block is connected to the next CNN block through a DC block, thereby providing a framework which iterates between intermediate de-aliasing and k-space correction. The first-order derivative of the forward pass in Eq. (4) with respect to the input can be easily expressed as $\frac{\partial f_{DC}}{\partial x_c} = F^H \Lambda F$, which enables the whole model to be trained in an end-to-end manner.

## 3.4 Vision Transformer, Swin Transformer, and Swin-Unet

The basic constituent of transformer networks is the self-attention mechanism. Given a sequence of vectorized image patch tokens $z_1, z_2, \ldots, z_n$ by $Z \in R^{n \times d}$, where $d$ is the dimension of the features, the main aim of self-attention is to capture relationships between all $n$ patches. This is done using three learnable weight matrices named Queries $(W^Q \in R^{d \times d_q})$, Keys $(W^K \in R^{d \times d_k})$, and Values $(W^K \in R^{d \times d_v})$. Given that $d_q = d_k$, the output $Z_{out} \in R^{n \times d_v}$ of a self-attention layer is given by:

$$Z_{out} = \text{softmax}\left(\frac{QK^T}{\sqrt{d_q}}\right)V \tag{5}$$

where $Q = ZW^Q$, $K = ZW^K$, and $V = ZW^V$.

Swin transformer, as discussed in Section 2, is a transformer variant which not only operates on local image patches but also models global correlations via a shifted window mechanism (See Fig. 1(c)). For a given sequence of vectorized image patch tokens $z_1, z_2, \ldots, z_n$, The operations within the Swin transformer block (See Fig. 3(a)) could be expressed as follows:

$$\hat{z}^l = \text{W-MSA}\left(\text{LN}(z^{l-1})\right) + z^{l-1} \tag{6}$$

$$z^l = \text{MLP}\left(\text{LN}(\hat{z}^l)\right) + \hat{z}^l \tag{7}$$

$$\hat{z}^{l+1} = \text{SW-MSA}\left(\text{LN}(z^l)\right) + z^l \tag{8}$$

$$z^{l+1} = \text{MLP}\left(\text{LN}(\hat{z}^{l+1})\right) + \hat{z}^{l+1} \tag{9}$$

where W-MSA and SW-MSA are window and shifted window multi-head self-attention, respectively. $\hat{z}^l$ and $z^l$ represents the output tokens of the W-MSA module and its preceding multi-layer perceptron (MLP) module, respectively. where $\hat{z}^{l+1}$ and $z^{l+1}$ represent the output tokens of the SW-MSA module and its preceding MLP module, respectively. In Fig 1(c), we illustrate how the shifted window mechanism works. During W-MSA, the green image patch interacts with all the other image patches within its window, and during SW-MSA, the receptive field of the green image patch has spanned across multiple windows in the image. This type of partitioning mechanism will enable the model to capture global correlations for an effective reconstruction. As discussed in Section 2, Swin-Unet, is an Unet-like pure transformer network which contains several Swin transformer blocks as can be seen in Fig. 3(c).

## 3.5 Point Spread Function

The PSF fundamentally describes the response to a point object and thus determines the quality of an imaging system as well as the captured image itself. Generally, for a linear, shift-invariant imaging system, the PSF governs the image of any known object via convolution as below (Robson et al., 1997):

$$I = I_0 * H \tag{10}$$

where $I$ is the underlying object, $H$ is the PSF of the imaging device, $I_0$ is the image captured from the said device, and $*$ is the convolution operation. In MR imaging, the k-space is sampled discretely via analog-to-digital converters utilizing square pulses, hence usually having a PSF of a $sinc$ function.

In accelerated MRI, the undersampled image, $I_{und}$ can be modelled using the PSF corresponding to the undersampling mask, $H_{samp}$ with respect to the fully sampled reference image, $I_{ref}$ as below (Tan and Zheng, 2005):

$$I_{und} = I_{ref} * H_{samp} \qquad (11)$$

where $H_{samp} = FM$ which is the Fourier transform of the sampling mask. Thus, the spatial spread of aliasing artifacts directly depends on $H_{samp}$, thereby on $M$. For instance, 1D undersampling in k-space will spread the aliasing artifacts row-wise in the subsequent aliased image as shown in Fig. 2.

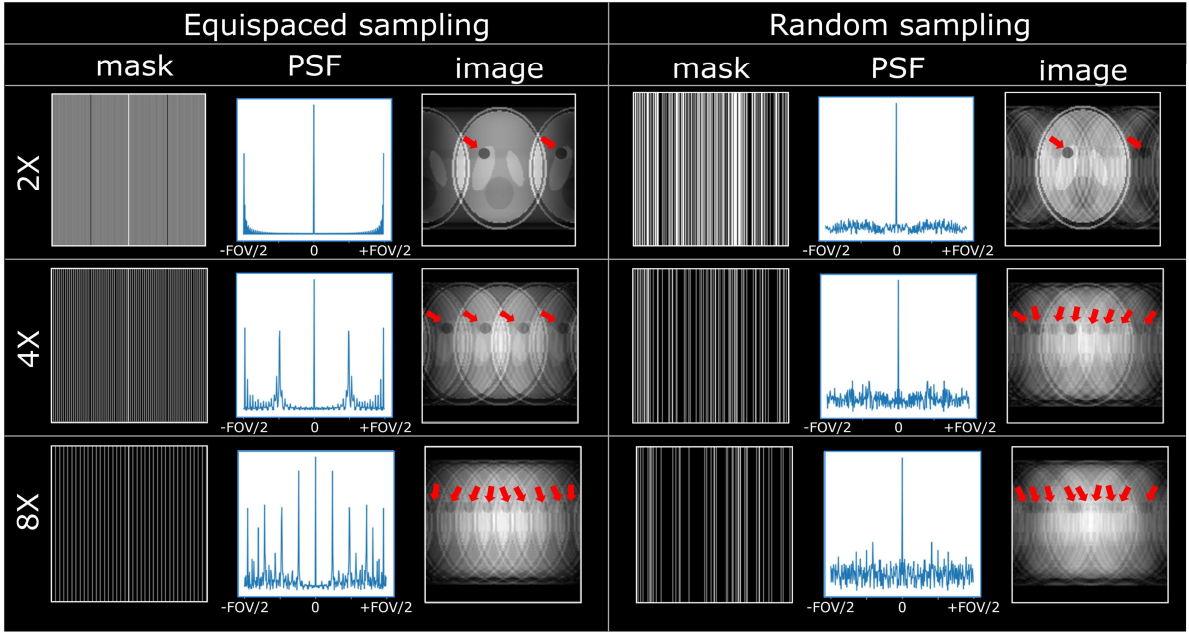

**Fig. 2.** Different sampling masks, their corresponding PSFs, and the resulting undersampled MR images with artifacts demonstrated on the Shepp–Logan phantom. Note the added dark grey spot which clearly depicts the coherence between the peaks of the PSF and the aliasing artifact pattern in the phantom image.

## 4  Methods

The overall schematic of the proposed McSTRA model is shown in Fig. 3. It consists of four main architectural components: the multi-branch feature extractor which captures image features corresponding to different spectral regions in k-space; the cascaded dual-channel Swin-Unet (DS) blocks interleaved by DC blocks and a discounted intermediate loss computation strategy; the reconstruction tail which enforces magnitude image reconstruction using a single-channel Swin-Unet (SS); the PSF-guided, instance-dependent, shared positional embedding generator which incorporates spatial spread in aliasing artifacts pattern. In this Section, we discuss each architectural component of McSTRA separately.

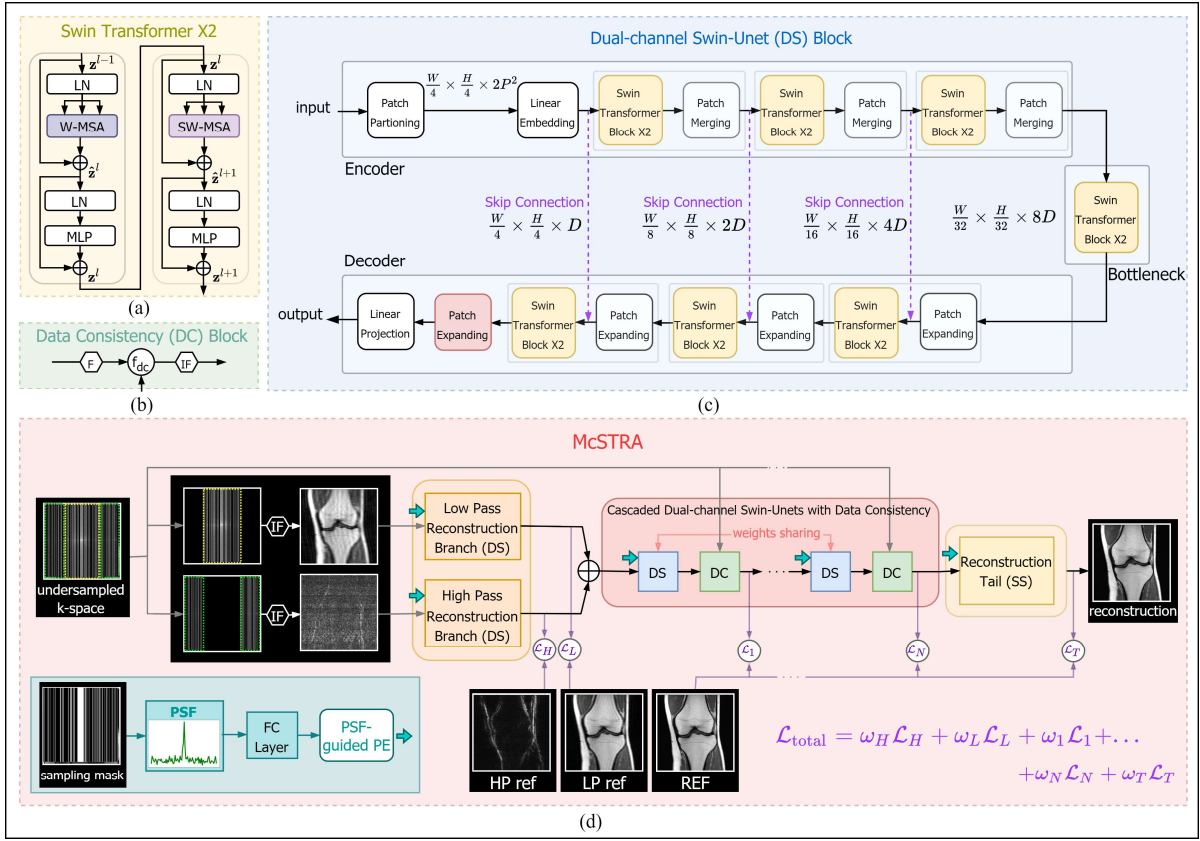

**Fig. 3.** Overview of McSTRA. (a) Swin transformer X2 block: W-MSA and SW-MSA (b) DC block (c) Swin-Unet architecture (d) Overall McSTRA model

### 4.1 Multi-branch Feature Extractor

Our proposed multi-branch feature extractor exploits the spectral properties in k-space by partitioning the k-space and extracting meaningful features separately. The multi-branch setup contains two branches. The low-pass branch is fed with only the low-frequency image features whereas the high-pass branch is fed with only the high-frequency image features as seen in Fig. 3(d) and Fig. A.1. The low- and high-frequency image features are obtained by partitioning the undersampled k-space into two further partitions and applying the Inverse Fourier transform. In order to partition the k-space, we choose twice as many lines as the retained center frequency lines (See Section 5.1 for the undersampling strategy) as the low-frequency partition while the complement is considered as the high-frequency partition. We found out that such a partitioning criterion effectively separates the anatomical and structural features from the edge and resolution features of the MR image as seen in Fig. A.1.

The intention here is to dedicate two distinguished sets of model parameters to learn the reconstruction of high- and low-frequency information of the image separately so that the high-frequency MR features will receive sufficient attention within the model. In the end, the outputs of the two branches will be fused together and passed down to the cascaded DS. As high- and low-pass branches, we utilize dual-channel Swin-Unet (DS) blocks which are fundamentally equivalent to the

Swin-Unet (Cao et al., 2021), but designed to handle complex MR data using two-channel input/output formation.

Given the undersampled k-space $\hat{y}$ as input, the output from the multi-branch setup, $x_{branch}$ can be expressed as below:

$$x_l = f_l(F^H(M_l \odot \hat{y})|\theta_l) \tag{12}$$

$$x_h = f_h(F^H(M_h \odot \hat{y})|\theta_h) \tag{13}$$

$$x_{branch} = x_l + x_h \tag{14}$$

where $x_l$ and $x_h$ are the outputs from the low- and high-pass branches, respectively. $f_l$ and $f_h$ represent the DS blocks corresponding to low-pass and high-pass branches, respectively. $M_l$ and $M_h$ are sampling masks which extract the center and outer regions of the undersampled k-space, respectively. In order to enforce additional regularization in terms of feature extraction, we compute the loss of individual branches and sum them up to contribute to the total loss of the model as below:

$$\mathcal{L}_{branch} = \alpha_l \cdot \mathcal{L}(\tilde{x}_l, x_l) + \alpha_h \cdot \mathcal{L}(\tilde{x}_h, x_h) \tag{15}$$

where $\tilde{x}_l = F^H(M_l \odot y)$ and $\tilde{x}_h = F^H(M_h \odot y)$ are the reference images corresponding to the low- and high-pass branches, respectively. We chose $\alpha_l = \alpha_h = \frac{1}{2}$.

## 4.2 Cascaded Dual-channel Swin-Unets

The cascaded DS blocks as seen in Fig. 3(d) are interleaved by DC blocks which perform correction in k-space (Fig. 3(b)) as discussed in Section 2. McSTRA consists of $N$ number of DS blocks interleaved by DC blocks forming an iterative cascade. Given the forward function of the $t^{th}$ DS block of the cascade as $f_{cas}$ with parameters $\{\theta_{cas}^t\}_{t=1}^N$, the undersampled k-space, $\hat{y}$ and the output from the multi-branch setup, $x_{branch}$, the output after the $t^{th}$ iteration can be expressed (following Eq. 4) as follows:

$$x_t = F^H \Lambda F f_{cas}(x_{t-1}|\theta_{cas}^t) + \frac{1}{1+\lambda} F^H \hat{y}, \quad t = 1, 2, \ldots, N \tag{16}$$

where $x_0 = x_{branch}$. Note that in our proposed framework, we share weights among the DS blocks for efficient computational memory usage, thus $\theta_{cas}^1 = \theta_{cas}^2 = \ldots = \theta_{cas}^N$. We compute a discounted loss along the cascade as below:

$$\mathcal{L}_{cas} = \sum_{t=1}^{N} \beta_t \cdot \mathcal{L}(\tilde{x}, x_t) \tag{17}$$

where $\tilde{x} = F^H y$ is the reference image. As seen in Eq. (17), not only the final reconstruction of the cascade, $x_N$, but also the intermediate reconstructions, $x_t$, contribute to $\mathcal{L}_{cas}$ multiplied by a discounted scale factor $\beta_t$. We set $\beta_t$ to increase linearly with $t$, i.e., $\beta_t = \frac{t}{\sum_{k=1}^{N} k}$ which enforces McSTRA to learn accurate MRI reconstructions incrementally. These intermediate loss computations also help to combat the vanishing gradients by additional regularization (Szegedy et al., 2015).

### 4.3 Reconstruction Tail

In McSTRA, the multi-branch, as well as the cascaded network, perform mappings from two-channel inputs to two-channel outputs facilitating real and imaginary parts of complex MR data and at the end of the cascaded network, image features have been sufficiently extracted as latent information. Hence, we designed the reconstruction tail to construct the final magnitude image. Thus, the reconstruction tail here acts as an optimized refining step to utilize the extracted information to reconstruct the optimal MR magnitude images. As the reconstruction tail architecture, we use a single-channel Swin-Unet (SS) to perform a magnitude image-to-magnitude image mapping. It is noted that the reconstruction tail can also be extended for the reconstruction of complex images. The forward function of the reconstruction tail can be expressed as below:

$$\boldsymbol{x}_{\text{tail}} = f_{\text{tail}}(\boldsymbol{x}_{\text{cas}}|\theta_{\text{tail}}) \tag{18}$$

Note that $\boldsymbol{x}_{\text{cas}}$ in Eq. (18) represents the magnitude image of the output from the cascaded DS. The tail loss is computed between $\boldsymbol{x}_{\text{tail}}$ and the magnitude reference image, $\tilde{x}$ as $\mathcal{L}_{\text{tail}} = \mathcal{L}(\tilde{x}, \boldsymbol{x}_{\text{tail}})$.

### 4.4 Loss Function

For a given batch during training, the total loss of McSTRA would be the weighted average of $\mathcal{L}_{\text{branch}}$, $\mathcal{L}_{\text{cas}}$, and $\mathcal{L}_{\text{tail}}$ computed as below:

$$\mathcal{L}_{\text{total}}(\theta) = \gamma_{\text{branch}} \cdot \mathcal{L}_{\text{branch}} + \gamma_{\text{cas}} \cdot \mathcal{L}_{\text{cas}} + \gamma_{\text{tail}} \cdot \mathcal{L}_{\text{tail}} \tag{19}$$

where $\theta$ represents all the trainable parameters of McSTRA. In our implementation, we chose $\gamma_{\text{branch}} = \gamma_{\text{cas}} = \gamma_{\text{tail}} = \frac{1}{3}$. Based on $\mathcal{L}_{\text{total}}$, back-propagation will be performed to update all the trainable parameters of the model including the positional embedding generation network discussed in Section 4.5.

### 4.5 Point Spread Function-guided Positional Embedding Generator

As discussed in Section 2, positional embeddings play a vital role in transformer networks in order to leverage the order of the sequence of tokens. As discussed in Section 3.5, the spatial spread of aliasing artifacts directly depends on the PSF. Thus, to guide the self-attention operations within the transformer layers, we proposed a semi-network which takes the PSF corresponding to the input image and generates a positional embedding for the respective input. This type of positional embedding could especially be beneficial in a random undersampling setting where each input is subjected to a unique undersampling pattern, hence consisting of a unique aliasing pattern compared with the rest of the inputs in the training dataset. Given the PSF-guided positional embedding generator network as $\mathcal{P}$, and the sampling mask for a given input as $\boldsymbol{M}$, the total positional embedding for a given input is computed as below:

$$\boldsymbol{E}_{\text{pos}} = \mathcal{P}(\boldsymbol{H}_{\text{samp}}|\theta_{\mathcal{P}}) + \boldsymbol{E}_{\text{abs}} \tag{20}$$

where $\boldsymbol{E}_{\text{abs}}$ is the conventional 1D positional embedding as in Dosovitskiy et al. (2021). As $\mathcal{P}$, we utilize a fully connected layer so that the resulting positional embedding for each pixel could be interpreted as weighted average of the underlying PSF. The parameters, $\theta_{\mathcal{P}}$ are updated using backpropagation on the total loss computation in Eq. (19). Moreover, we share the positional embedding, $\boldsymbol{E}_{\text{pos}}$ across all Swin-Unets in McSTRA.

### 4.6 Implementation

For the self-attention layers in the multi-branch setup, the cascaded DS, and the reconstruction tail, we utilize an embedding dimension of 48, 96, and 48 respectively. As the loss, $\mathcal{L}$, for all computations we utilized $\ell_1$-norm-based loss (Zbontar et al., 2019). We implement five iterations within the cascade, i.e., $N = 5$ following similar formulation as in Schlemper et al. (2018). McSTRA was trained on an NVIDIA A40 GPU for 50 epochs in a batch size of 8 using RMSProp optimizer with a learning rate of $1e^{-4}$.

## 5 Experiments and Results

In this Section, we present the empirical results of the experiments that we conducted in order to assess the validity of McSTRA as a compatible solution for accelerated MRI Reconstruction.

### 5.1 Experimental Setting

**Datasets and Preprocessing**. For the training and validation of the McSTRA, the fastMRI knee and brain datasets (https://fastmri.org/dataset/) were utilized. The knee dataset consists of fully sampled knee MR k-space data obtained on 1.5 and 3.0 Tesla magnets. The raw complex k-space data include coronal proton density-weighted data with and without fat suppression (PDFS and PD respectively). The exact number of MR volumes and 2D slices utilized for training and validation are given in Table 1. All models (McSTRA and other comparative methods) were trained on PD and PDFS data together. The intensity values of the knee data were scaled in the range [0,1] to avoid vanishing gradients which could cause training to fail.

Table 1. The numbers of volumes and 2D slices of the fastMRI knee dataset.

| Sequence | Training | Validation | Total |
|---|---|---|---|
| PD | 484 [17286] | 100 [3592] | 584 [20878] |
| PDFS | 489 [17456] | 99 [3543] | 588 [20999] |
| Total | 973 [34742] | 199 [7135] | 1172 [41877] |

**Undersampling.** The input to the model is undersampled complex k-space data which were obtained by retrospectively masking k-space lines in the phase encoding direction. Following a similar undersampling procedure to Zbontar et al. (2019), during training, we applied the same undersampling mask to all the slices in a given volume with a random four-fold acceleration while retaining 8% of the center frequencies.

**Comparative Methods.** In order to assess the competitiveness of the proposed model, we compared its reconstruction accuracy with several other state-of-the-art DL-based MRI reconstruction methods, i.e. the fastMRI U-Net baseline (Zbontar et al., 2019), a deep cascade of CNNs (D5C5) (Schlemper et al., 2018), MRI dual-domain reconstruction network (MD-Recon-Net) (Ran et al., 2021), vision transformer (ViT) (Lin and Heckel, 2021), and Swin transformer for fast MRI (SwinMR) (Huang et al., 2022). For the fairness of comparison, we sustained the original hyperparameters and the training details for these models following their original literature. We implemented the single-channel setting of SwinMR as the original work was proposed for multi-channel reconstruction.

**Evaluation Metrics.** For quantitative performance evaluation purposes, we utilize three commonly used metrics, i.e., normalized mean square error (NMSE), peak signal-to-noise ratio (PSNR), and structural similarity (SSIM) following formal definitions in Zbontar et al. (2019). We present the volume-wise mean as well as the standard error of means (e.g., Figures 4, 6, 8, 9, 10, 12, 14, 15, and 16) with respect to the above metrics in order to better comprehend the confidence level of quantitative scores. The tabulation of scores can be found in Appendix Section A.2 as Table A.1-A.9.

## 5.2 Reconstruction Performance

Fig. 4 summarizes the quantitative reconstruction performance of different methods on the fastMRI knee validation dataset where we present the results for PD and PDFS data separately and together (Overall). Here the undersampling utilized is the same as the training setting, i.e., random four-fold acceleration while retaining 8% of the center frequencies. It can be seen that McSTRA outperforms all the other comparative methods in terms of NMSE, PSNR, and SSIM. To test whether these improvements were statistically significant, a non-parametric Wilcoxon rank test was utilized and $p < 0.05$ was considered statistically significant. McSTRA performed significantly better than U-Net, MD-Recon-Net, and SwinMR with $p < 0.05$. When compared with D5C5, the improvement in McSTRA is not statistically significant but a consistent trend in improvements was still observed. When compared with ViT, McSTRA had significantly better performance in NMSE and PSNR. It was also observed that generally, PDFS reconstructions show inferior performance than PD reconstructions mainly due to the inherent noise in PD acquisitions of the fastMRI dataset. Fig. 5 shows an example PD knee slice reconstruction from the fastMRI validation dataset. It demonstrates the superiority of McSTRA in reconstructing fine anatomical details unlike the other CNN-based methods such as U-Net and MD-Recon-Net which produce blurry outputs where fine anatomical details are present. It also demonstrates McSTRA's superior capability in reconstructing anatomical boundary features (see the red and yellow patches in Fig. 5) and textural details in bone (see the blue patch in Fig. 5) which specifically is a consequence of the multi-branch setup which enforces sufficient attention within the overall network to the high-frequency MR features which often contains resolution and edge information.

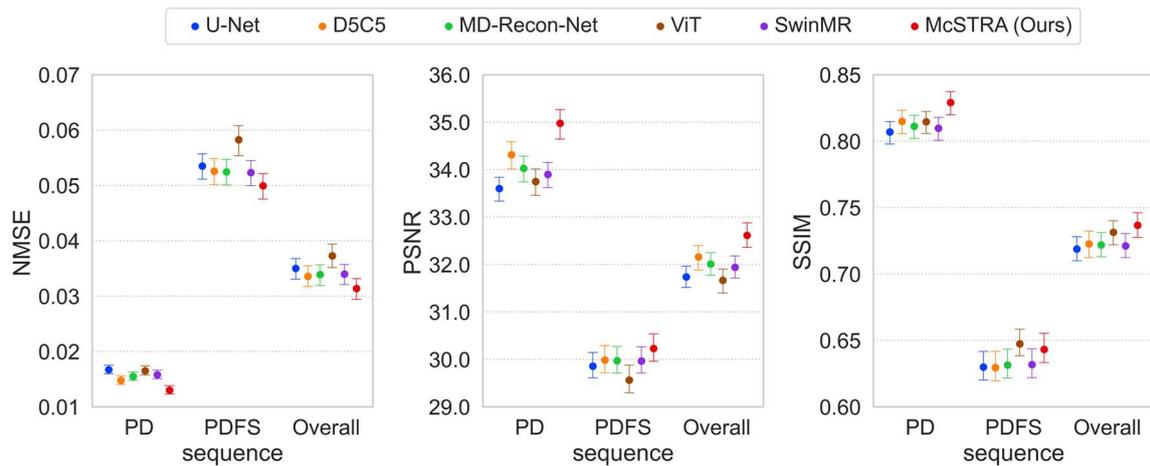

**Fig. 4.** Quantitative performance comparison of McSTRA under random four-fold acceleration while retaining 8% of the center frequencies on the fastMRI knee validation dataset.

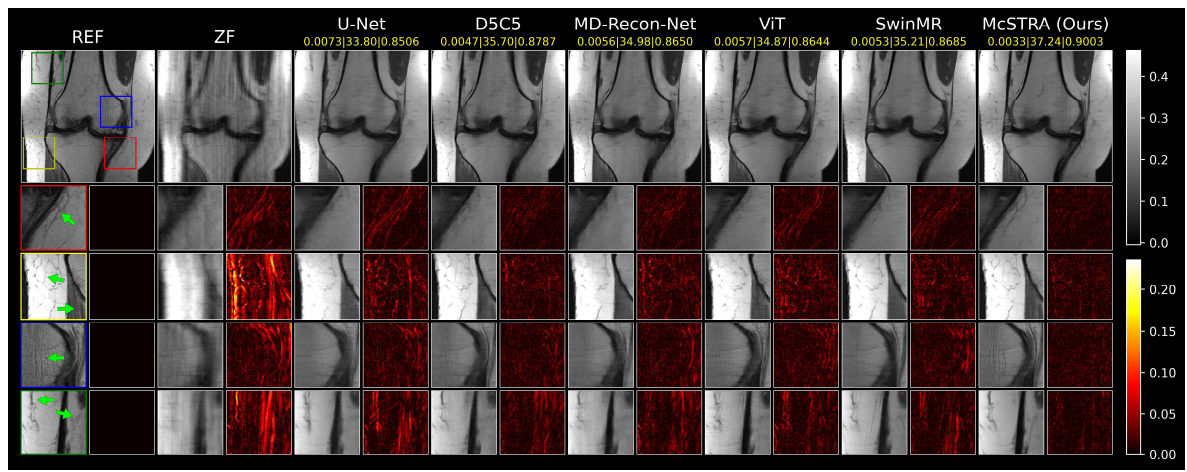

**Fig. 5.** A PD knee slice reconstruction from the fastMRI validation dataset using different methods under four-fold acceleration while retaining 8% of the center frequencies (Row 1). The NMSE, PSNR, and SSIM scores for each method for the reconstructed slice are displayed on the top (in yellow), respectively. Four regions of interest (red, yellow, blue, and green patches) and their corresponding error images are enlarged side-by-side in Rows 2-5. We direct readers' attention to the green arrows shown in the reference image which indicate fine anatomical features.

## 5.3 Robustness to Noise

In this experiment, we evaluated the robustness of McSTRA to noisy MR acquisitions. We introduced complex additive Gaussian noise to the input k-space measurements (baseline), thereby, simulating signal-to-noise ratio (SNR) levels of 50, 20, 15, 10, 5, and 0 dB, and observed the reconstruction performance of McSTRA and other comparative methods on the fastMRI knee validation dataset. Here the undersampling utilized is the same as the training setting. We present the results for PD and PDFS data together in Fig. 6 which depicts the quantitative reconstruction performance of different methods at different SNRs. It can be seen that McSTRA outperforms all methods at all noise levels. Note that the results for U-Net, ViT, and SwinMR at 0 dB are not shown due to the large deterioration of those methods at 0 dB. It is also seen that the performance gap between McSTRA and other methods with respect to NMSE, PSNR, and SSIM increases with SNR which is an indication of how comparatively well McSTRA oversees noisy data. To test whether these improvements were statistically significant, a non-parametric Wilcoxon rank test was utilized. The p-value significance level was $p < 0.05$ between McSTRA and all the other methods at all noise levels tested above the baseline. In Fig. 7, we visualize reconstructions of a PD knee slice from the fastMRI validation dataset under different SNR levels. Generally, for SNR levels above 15 dB, all methods are able to reconstruct most of the fine features in bone and tissue. But, for SNR levels below 15 dB, it can be seen that CNN-based methods such as D5C5 and Md-Recon-Net drastically fail to reconstruct fine features. It is also worthwhile to note, that even when the SNR is 0 dB where noise power is equal to signal power, McSTRA is still able to reconstruct images with reasonable contrast whereas MD-Recon-Net and ViT demonstrate significant errors.

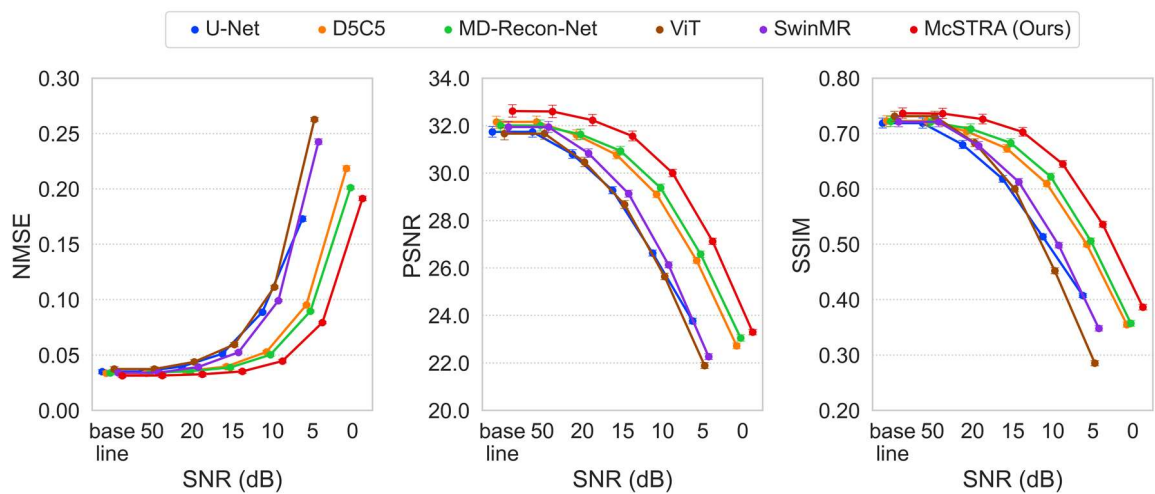

**Fig. 6.** Quantitative reconstruction performance of McSTRA at different SNRs under random four-fold acceleration while retaining 8% of the center frequencies on the fastMRI knee validation dataset (PD and PDFS together).

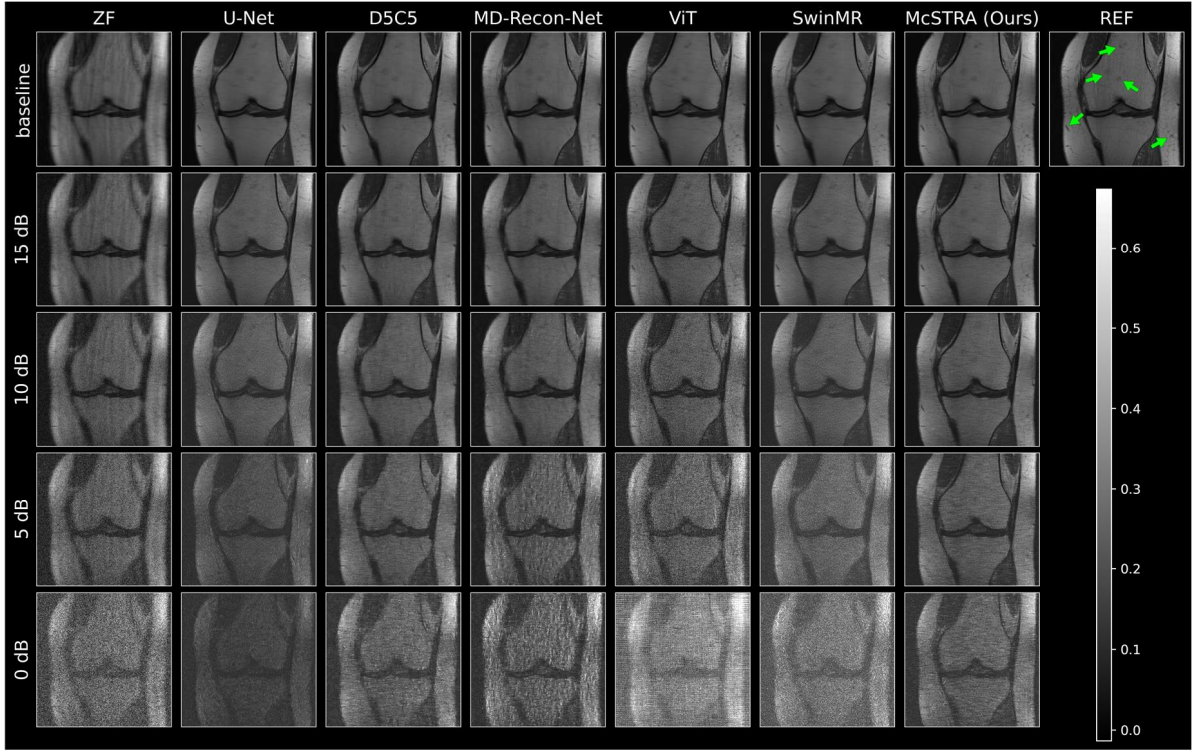

**Fig. 7.** A PD knee slice reconstruction from the fastMRI validation dataset under different SNR levels. We direct readers' attention to the green arrows in the reference image which indicate fine anatomical features which McSTRA shows superior capability of preserving even at high noise levels.

### 5.4 Sampling Changes during Inference

In this experiment, we studied the robustness of McSTRA to different sampling settings with respect to acceleration factor and mask type.

### 5.4.1 Changes in Acceleration Factor

Here we increased the acceleration factor from 4 to 12 in steps of two and assessed the reconstruction performance of McSTRA and other comparative methods on the fastMRI knee validation dataset. We present the results for PD and PDFS data together in Fig. 8. McSTRA's performance in terms of NMSE, PSNR, and SSIM drops gradually with the acceleration factor, nevertheless, it outperforms all other methods with respect to NMSE and PSNR under all acceleration factors. Further, a Wilcoxon rank test showed a p-value significance level of $p < 0.05$ between McSTRA and all the other methods under all acceleration factors with respect to all metrics except when compared with ViT using SSIM for the six-fold acceleration. Visual comparisons are shown in Fig. A.2. In terms of SSIM scores, McSTRA's performance is lower than ViT at acceleration factors of 8, 10, and 12.

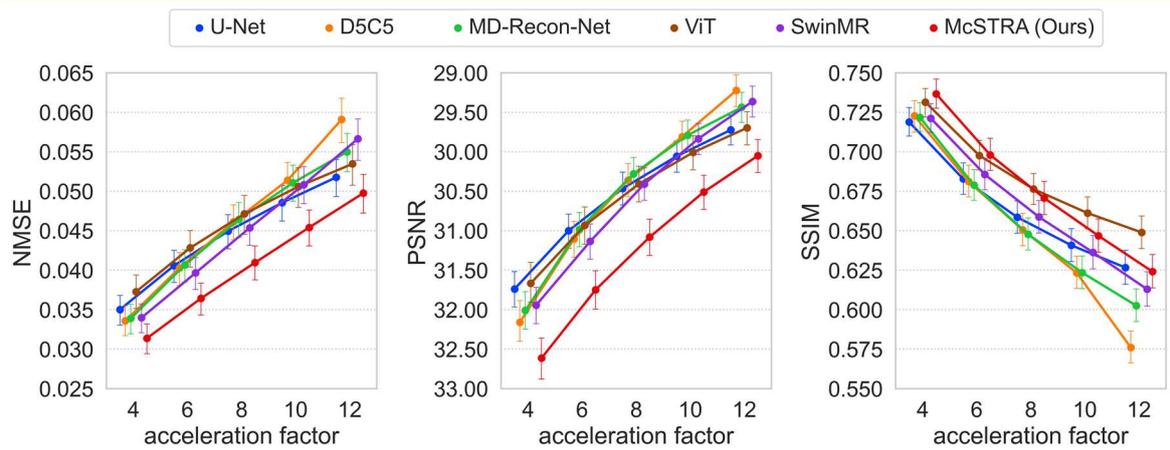

**Fig. 8.** Quantitative reconstruction performance of McSTRA at different acceleration factors while retaining 8% of the center frequencies on the fastMRI knee validation dataset (PD and PDFS together).

### 5.4.2 Changes in Mask Type

In this experiment, we assessed the reconstruction performance of McSTRA with equispaced k-space sampling on the fastMRI knee validation dataset. We present the results for PD and PDFS data together in Fig. 9 which demonstrates that McSTRA is able to retain superior performance in terms of NMSE, PSNR, and SSIM. Note that the change in performance of McSTRA is minute when compared to the results of the random sampling scenario in Section 5.2 (Fig. 4) which is an indication of the robustness of McSTRA to the undersampling protocol. Also, a Wilcoxon rank test showed a p-value significance level of $p < 0.05$ between McSTRA and all the other methods under equispaced sampling protocols.

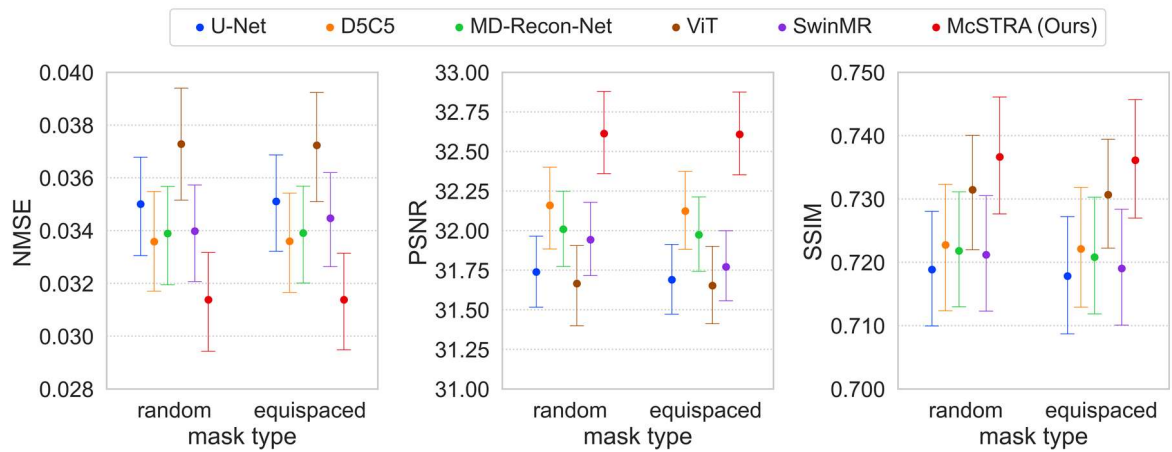

**Fig. 9.** Quantitative reconstruction performance of McSTRA for equispaced sampling protocol in k-space under random four-fold acceleration while retaining 8% of the center frequencies on the fastMRI knee validation dataset (PD and PDFS together).

## 5.5 Inference on Brain MR Data

In this experiment, we evaluated the performance of McSTRA on brain MR data. We utilized the models that were trained on the fastMRI knee dataset to reconstruct brain slices of the fastMRI brain validation dataset which was also retrospectively undersampled with a random four-fold acceleration while retaining 8% of the center frequencies. The fastMRI brain validation dataset consists of more than 1378 fully sampled knee MR k-space volumes (21842 slices) obtained on 3 and 1.5 Tesla magnets. The raw k-space dataset includes axial T1-weighted, T2-weighted, and FLAIR images. Some of the T1 weighted acquisitions included admissions of a contrast agent (T1POST). We present the results for T1, T1POST, T2, and FLAIR reconstructions separately and together (Overall) in Fig. 10 where it can be observed that McSTRA demonstrates superior performance across all image sequences in terms of NMSE, PSNR, and SSIM. Note that, McSTRA achieves very high PSNR and SSIM (e.g., T1POST: PSNR of 38.68 ± 0.11 and SSIM of 0.9536 ± 0.0008) despite the fact that the model was trained on PD and PDFS knee data.

Fig. 11 shows an example reconstruction of a T1 brain slice from the fastMRI brain validation dataset which further illustrates that McSTRA is able to resolve most of the aliasing artifacts while preserving comparatively high contrast between grey matter and white matter without model re-training. Here, CNN-based methods such as U-Net as well as CNN-transformer hybrid methods such as the SwinMR drastically fail to oversee aliasing artifacts, hence producing corrupted images with low quality. We believe that a number of reasons have influenced McSTRA to produce robust reconstructions when inferred on a dataset of a different anatomical region. One reason is that PSF-guided positional embeddings that guide the self-attention mechanism by generating a unique positional embedding for each input based on the PSF of the sampling mask, which is independent of image content. The other reason to obtain high-quality results compared to the other methods is the k-space partitioning mechanism in McSTRA which gives considerable attention to the high-resolution features of the MR image. Finally, the global feature extraction from the transformer-based backbone utilized in McSTRA influenced the robustness as the underlying self-attention mechanism generates weights by considering the similarities within the image itself for a long-range feature aggregation, unlike the local CNN-based feature aggregation in each layer which heavily correlates neighboring anatomical features, and limits their generalization towards different anatomical characteristics beyond the training dataset.

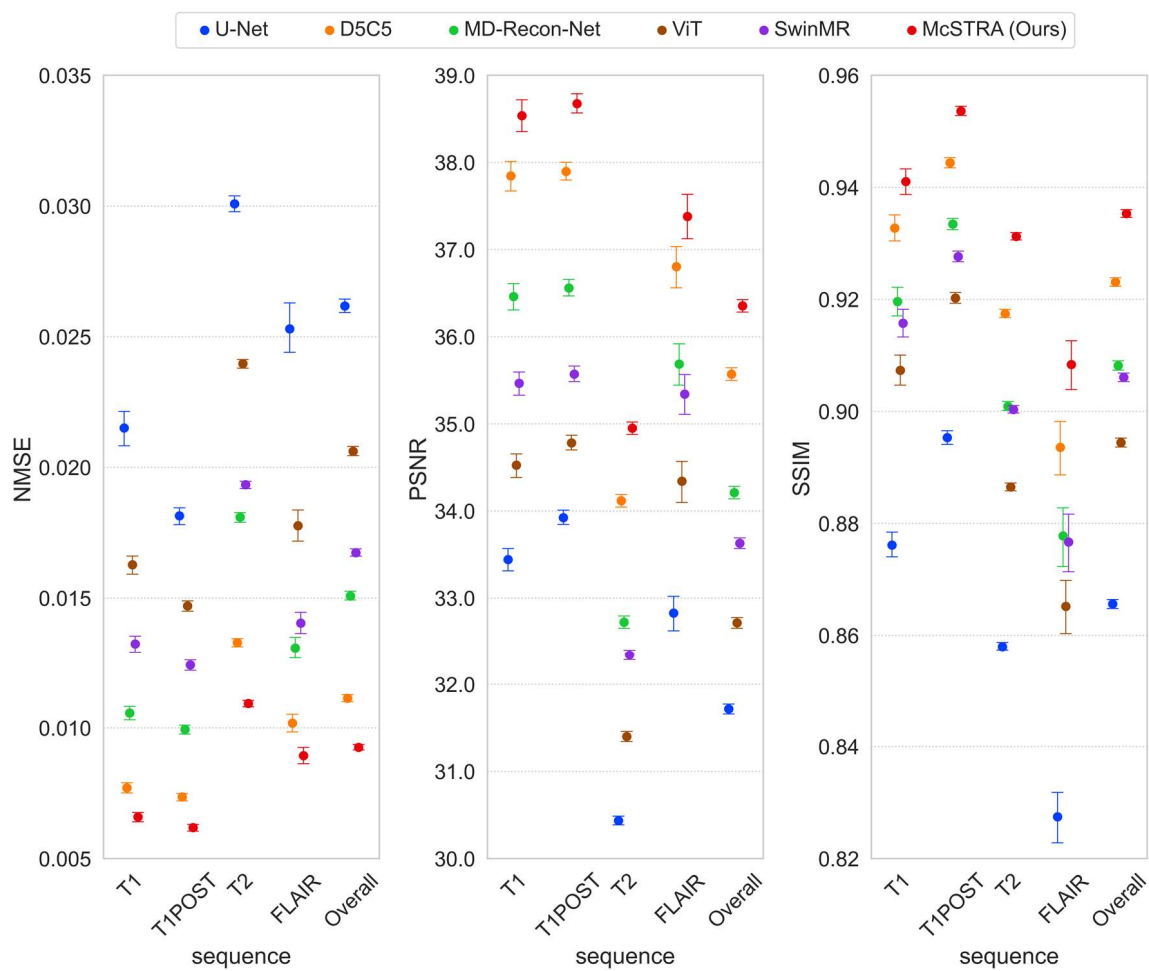

**Fig. 10.** Quantitative reconstruction performance of McSTRA when inferred on fastMRI brain validation dataset under random four-fold acceleration while retaining 8% of the center frequencies (best performance for each sequence in boldface).

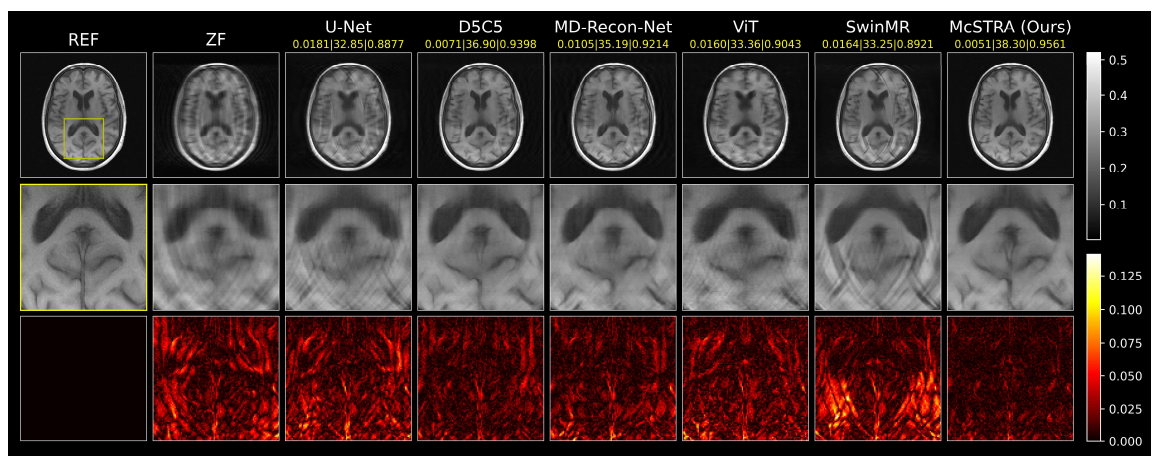

**Fig. 11.** Inference on a T1 brain slice from the fastMRI brain validation dataset under random four-fold acceleration while retaining 8% of the center frequencies (Row 1). The NMSE, PSNR, and SSIM scores for each method for the reconstructed slice are displayed on the top (in yellow). The region regions of interest (yellow patch) and their corresponding error image are enlarged in Rows 2 and 3, respectively.

## 5.6 Robustness to Abnormalities

Irrespective of the drastic improvement in accelerated MRI using DL, false-negative reconstructions are still a concerning issue (Antun et al., 2020). During the presentation of results of the fastMRI challenge 2019 and 2020 (https://slideslive.com/38922093/medical-imagingmeets-neurips-4), it was revealed that even high-performing DL models are unable to reconstruct small abnormalities such as meniscal tear and subchondral osteophytes. Further, the results of the study conducted by Recht et al. (2020) demonstrate that it is often difficult to reconstruct small structures near the knee joint which could be clinically relevant not only for diagnosis but also for the treatment of lesions. In support of battling such adversarial attacks on image reconstruction, Calivá et al. (2021) manually annotated MRI slices of the fastMRI knee dataset which contain lesions.

In this experiment, we evaluated our trained models on those MRI volumes identified by Calivá et al. (2021) from the fastMRI knee validation dataset (which includes both PD and PDFS knee data) in order to determine McSTRA's ability to battle abnormalities in anatomical structures. Fig. 12 presents quantitative scores categorized based on the location of the knee where the lesion is present, i.e., bone marrow, cartilage, meniscus, or cyst. It is seen that McSTRA is able to achieve high PSNR and SSIM, especially when the lesion exists in the bone marrow or cartilage. The performance is comparatively high even when the lesion exists in the cyst compared to other methods like ViT and CNN-based methods like U-Net which show inferior capabilities for reconstructing abnormalities in terms of NMSE, PSNR, and SSIM. Fig. 13 shows an example reconstruction of a PDFS knee slice from the fastMRI knee validation dataset which was identified to contain a lesion in the bone marrow region. This visual comparison further illustrates McSTRA's superior capabilities of recovering abnormalities and overseeing out-of-distribution data. Compared with the other methods, McSTRA shows strong capabilities of reducing artifacts specifically in the region of the lesion.

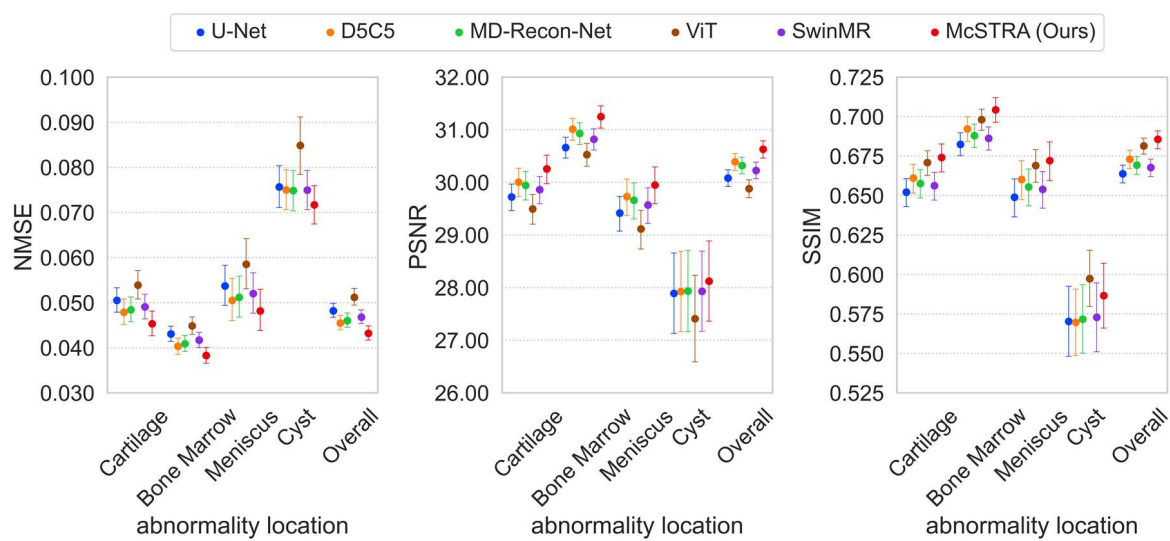

**Fig. 12**. Quantitative performance comparison of McSTRA under random four-fold acceleration while retaining 8% of the center frequencies on the fastMRI knee validation volumes which was identified to contain lesions by Calivá et al. (2021).

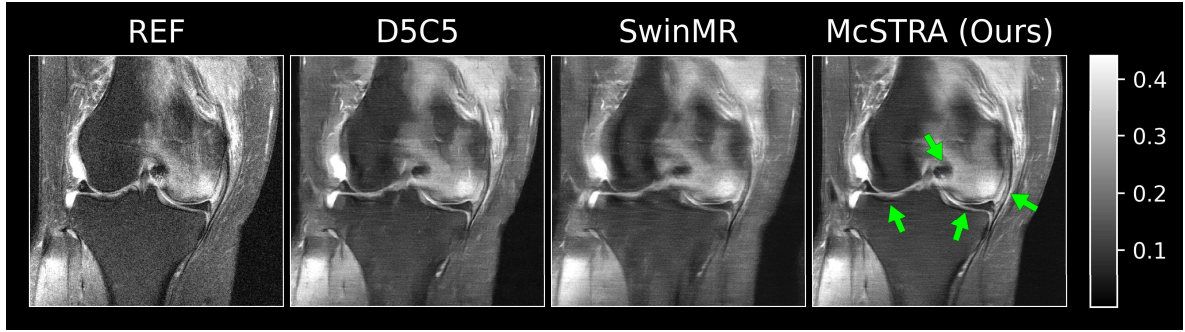

**Fig. 13.** A PDFS knee slice reconstruction from the fastMRI validation dataset using different methods under random six-fold acceleration while retaining 5.33% of the center frequencies, which contains a lesion in the bone marrow identified by Calivá et al. (2021). We direct readers' attention to the green arrows shown in McSTRA's reconstruction which indicate fine features near the location of the lesion.

### 5.7 Ablation Studies

Here, we assessed the effectiveness of each constituent component of McSTRA and their contribution to the overall improved performance of McSTRA.

#### 5.7.1 Effectiveness of the Multi-branch Feature Extraction

Here we performed experiments with and without the multi-branch setup on the fastMRI knee validation dataset. We present the results for PD and PDFS data together in Fig. 14 as Configuration A where it can be seen that the multi-branch feature extraction setup which focuses on complementary k-space regions improves the performance of McSTRA for accurate MRI reconstruction in terms of NMSE, PSNR, and SSIM.

For further investigation, we performed an experiment to keep the number of parameters of the model unchanged and we compared the effectiveness of the proposed k-space partitioning mechanisms. Here, we utilized the same McSTRA network architecture but fed the MR image corresponding to the whole undersampled k-space to the low-pass as well as the high-pass branches without partitioning and the loss computation at each branch was modified to reconstruct the full image. The results are shown in Fig. 14 as Configuration B. By comparing the scores of Configuration B with Configuration F (McSTRA), it is observed that modelling the low- and high-frequency k-space components separately considerably improved the performance.

Also, we implemented an alternative k-space partitioning criterion where the low-frequency partition is a square region in the center of the k-space as in Faris et al. (2021). We observed that such a partitioning criterion produces deteriorated quantitative scores compared to that of McSTRA. The results are shown in Fig. 14 as Configuration E. By comparing the scores of Configuration E with Configuration F (McSTRA), it is observed that including some of the high-frequency components in the low-frequency branch as in McSTRA reinforces accurate reconstructions.

Also, we analyzed the sensitivity of reconstruction performance to the loss coefficient of the low-pass and high-pass branches, i.e., $\alpha_l$ and $\alpha_h$. Results are shown in Fig. 15 where it is evident that $\alpha_l/\alpha_h = 1$ yields optimal NMSE, PSNR, and SSIM scores. It was also noted that increasing as well as decreasing $\alpha_h/\alpha_l$ will deteriorate the reconstruction performance, yet, decreasing $\alpha_h/\alpha_l$ yields comparatively better performance than increasing $\alpha_h/\alpha_l$.

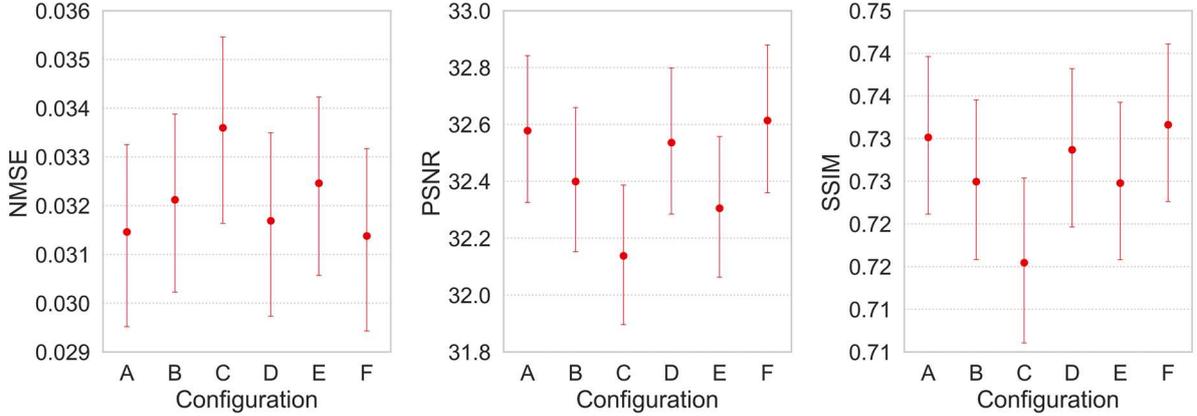

**Fig. 14.** Quantitative reconstruction performance of McSTRA with and without its constituent architectural components under random four-fold acceleration while retaining 8% of the center frequencies on the fastMRI knee validation dataset. Configuration A: Without the multi-branch setup. Configuration B: Each branch of the multi-branch feature extractor is fed with the same unpartitioned k-space and the loss computation at each branch is based on the full k-space. Configuration C: Reconstruction tail replaced by a Dual-channel Swin-Unet which performs a complex-valued mapping. Configuration D: Without the PSF-guided positional embeddings. Configuration E: Partitioning the k-space in such a way that the low-frequency partition is a square region in the center of the k-space. Configuration F (McSTRA): The overall McSTRA framework with all the constituent components.

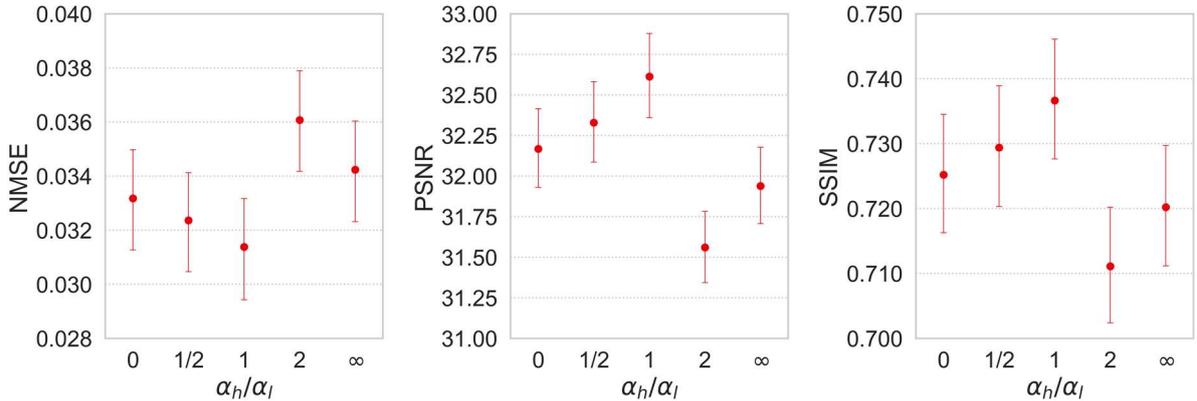

**Fig. 15.** Reconstruction performance of McSTRA with varying $\alpha_h/\alpha_l$.

### 5.7.2 Effectiveness of the Cascaded Network

Here we assessed the effectiveness of the cascaded network of Swin-Unets and the intermediate loss computations. We trained and evaluated McSTRA with and without intermediate loss computations after each iteration. We present the results for PD and PDFS data together on the fastMRI knee validation dataset in Fig. 16 where the performance gradually increases through the cascade for the model with the intermediate loss computations resulting in a better final performance in terms of NMSE, PSNR, and SSIM. Even the model without the intermediate loss computations shows increments in performance through the cascade, but those increments are not gradual resulting in lower final performance in terms of NMSE, PSNR, and SSIM. Fig. 16 also shows that the difference in the performance of the two models converges with iterations which is an indication that the model trained without intermediate loss computations reconstructs more accurate images during the later iterations and not during the earlier iterations. This signifies that the backpropagation of the overall loss does not propagate to the first several iterations sufficiently. In contrast, the model with intermediate loss computations at each iteration enforces accurate reconstruction after each iteration which ultimately leads to better final performance.

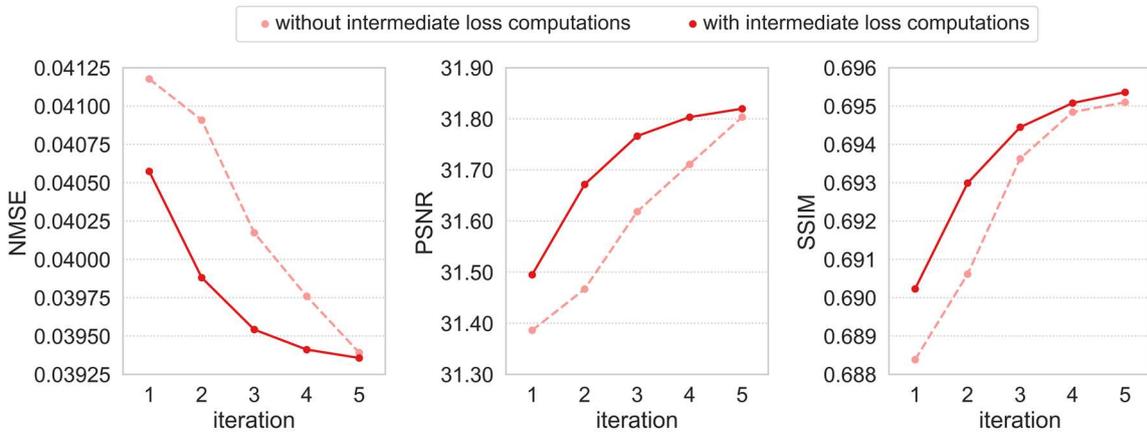

**Fig. 16.** Comparative reconstruction performance of the models trained with and without the intermediate loss computations under random four-fold acceleration while retaining 8% of the center frequencies on the fastMRI knee validation dataset.

### 5.7.3 Effectiveness of the Reconstruction Tail

Both the Multi-branch Feature Extractor as well as the cascaded network operate in the complex domain in the form of two-channel modules catering to real and complex counterparts of the k-space as well as the image data. For the reconstruction tail, we deployed a single-channel Swin-Unet to perform a magnitude image-to-magnitude image mapping. In this ablation study, we show that such an approach performs quantitatively better than enforcing a complex-valued mapping at the final stage of McSTRA. For a fair comparison, we retained all the other elements of McSTRA unchanged but the reconstruction tail comprised a dual-channel Swin-Unet to cater to complex-valued (real and imaginary components) data. The results are reported in Fig. 14 as Configuration C. By comparing the scores of Configuration C with Configuration F (McSTRA), it can be seen that a magnitude image mapping produces better reconstruction accuracy.

### 5.7.4 Effectiveness of the PSF-guided Positional Embeddings

This experiment shows how PSF-guided positional embeddings provide information for McSTRA during the learning process. In Fig. 17, we illustrate the cosine similarity correspondence amongst the PSF-guided positional embedding patches. It can be observed that the positional embeddings generated following Eq. (20) learn positional information quite accurately. We illustrate several cosine similarity maps corresponding to different regions of an MR slice selected from the fastMRI knee validation dataset. It is observed that these positional embeddings have been able to direct attention to the image patches with similar anatomical information. The effectiveness of the positional embeddings on quantitative image reconstruction performance was assessed by implementing McSTRA without the PSF guided positional embedding and is shown in Fig. 14 as Configuration D. Thereby, it is evident that the PSF guided positional embedding generation contributes to the improvement of reconstruction accuracy.

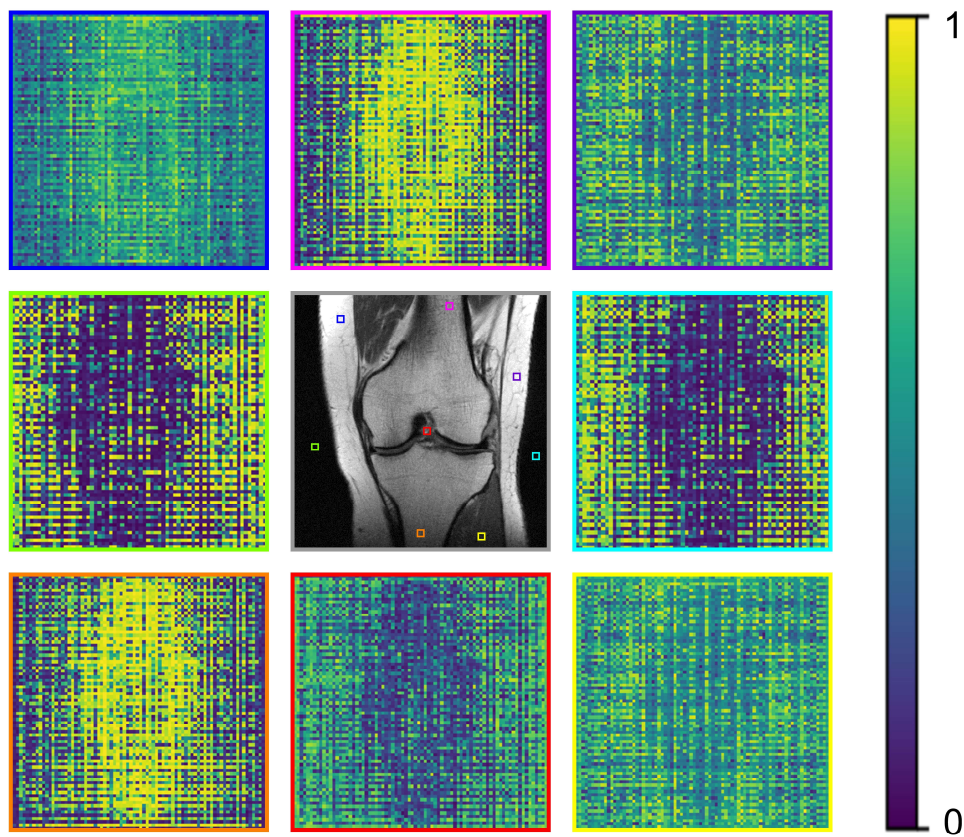

**Fig. 17.** Cosine similarity between positional embeddings corresponding to some selected patches (demonstrated using distinct colors) of the image and the positional embeddings corresponding to the rest of the patches of the center image.

### 5.8 Convergence Analysis

In Fig. 18, we compare the convergence of the variants of McSTRA presented in previous Sections by providing training and validation curves for the fastMRI knee dataset. It is clearly seen that the final

McSTRA has a smoother training loss curve compared to the other variants. Thus, it is evident that the incorporation of the multi-branch feature extraction and the PSF-guided positional embedding generation has clearly benefited McSTRA's convergence. Note that we exclude Configuration B from Fig. 18 due to fluctuations in convergence resulting in poor representation.

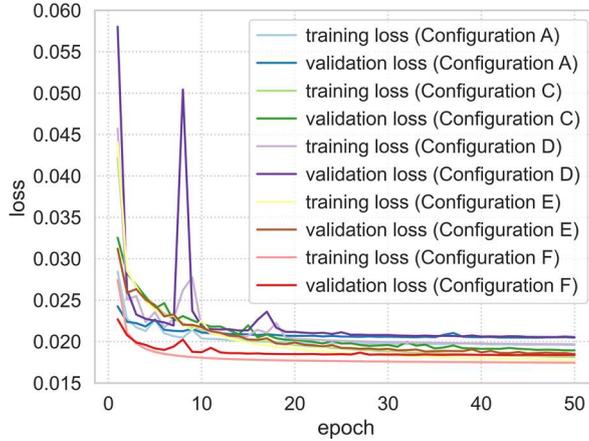

**Fig. 18.** Training and validation curves of McSTRA and its variants (presented in Fig. 14) for the fastMRI knee dataset.

## 5.9 Model size and Inference time

To assess computational costs, we provide the number of trainable parameters and inference time for each model that we trained in Table 2. The inference was performed on an NVIDIA GeForce RTX 3080 Laptop GPU. We observed that the CNN-based methods generally consisted of less trainable parameters and shorter inference times as a result of the inherent convolution operations and in contrast, the transformer-based models consist of more trainable parameters as they comprise fully connected layers. Considering the transformer-based networks, ViT has ≈1.5 times the trainable parameters as McSTRA. SwinMR consists of a smaller number of trainable parameters compared to McSTRA, but, the inference time is ≈3 times that of McSTRA due to the cascaded Swin Transformer layers in SwinMR's architecture. Overall, we could conclude that McSTRA has a moderate model size in terms of trainable parameters as well as a moderate inference time compared with the other methods.

**Table 2.** The number of trainable parameters and the inference time (per single MR slice of 320X320) of the reconstruction algorithms.

| Method | No. of trainable parameters | Inference time (seconds) |
|---|---|---|
| U-Net | 7756097 | 0.009 |
| D5C5 | 565770 | 0.131 |
| MD-Recon-Net | 289319 | 0.026 |
| ViT | 71821924 | 0.010 |
| SwinMR | 2223889 | 1.808 |
| McSTRA (Ours) | 48421488 | 0.501 |

# 6   Discussion and Conclusions

In this work, we have introduced a transformer-based DL model which incorporates MR physics for accelerated MRI reconstruction. The overall model is capable of reconstructing high-quality MR images from raw complex MR k-space data outperforming state-of-the-art DL models for MRI reconstruction. In the reported experiments, McSTRA shows superior robustness to adverse conditions such as higher accelerations, noisy data, different undersampling protocols, out-of-distribution data, and abnormalities in anatomy. In the Ablations studies, we illustrate the effect of each constituent component of McSTRA which confirms the suitability of the underlying rationale behind the conceptual design. McSTRA shows smoother convergence in training compared to its counterpart models without multi-branch feature extraction and the PSF-guided positional embedding generation which further illustrates the effectiveness of these architectural components.

One of the key observations from our experiments was the capability of McSTRA to reconstruct fine anatomical features. For instance, the results in Fig. 5 illustrate how edge features in tissue and textural information in bone are recovered during reconstruction by McSTRA whereas the other methods drastically fail to do so. Also, our experiments highlight the capability of McSTRA to oversee noise in k-space measurements. Such improvements are from the global attention mechanism employed in the transformer models. As seen in Fig. 7, even at extreme SNR levels such as 0 dB, McSTRA is able to reconstruct the finest anatomical structures in the knee unlike the CNN-based models such as U-Net and D5C5 which fails to eliminate noise. Furthermore, McSTRA's reconstruction performance with respect to NMSE and PSNR stands out from all other CNN-based models and ViT even at exceedingly high accelerations factors such as 10 and 12. McSTRA's robustness to sampling protocol is further established by the results in Fig. 9 where NMSE, PSNR, and SSIM deteriorates only by a small amount when inferred under an equispaced undersampling setting compared to random undersampling. We believe these outcomes are an indication of the potential of physics-based transformer models to be utilized in the general clinical setting where generalizability and robustness are key requirements.

The inference on the brain dataset further reinforces this rationale where the model that was trained on PD and PDFS knee data displays exceptional de-aliasing capabilities and comparatively high tissue contrast when inferred on T1, T2, and FLAIR brain data. As seen in Fig. 11, CNN-based models such U-Net as well as CNN-transformer hybrid models such as SwinMR fail drastically to generalize on different MR sequences and anatomy during inference. McSTRA's ability to oversee out-of-distribution data is further confirmed by the experiments that were conducted on the knee dataset with lesions. The quantitative scores in Fig. 12 clearly demonstrate the superiority of McSTRA in handling abnormalities in the knee which could be extremely beneficial in the clinical setting for diagnosis as well as treatment monitoring.

Despite many strengths, we identify several limitations of the proposed method. One limitation of the proposed model architecture is that it is designed for the single-channel MR acquisition setting whereas many MRI data are acquired in a multi-channel setting. However, this work demonstrates the fundamental principle of McSTRA, and a multi-channel setup can be readily extended based on the current model. In addition, McSTRA has a transformer-based backbone constructed by fully-connected layers which do not support varying input sizes. This limitation could be overcome by resizing the images before input into the model. Although sampled k-space points need to be transformed correspondingly for modelling data consistency due to resizing data in image space, it is a useful strategy in recent literature (Ottesen et al., 2022) and it also has an added advantage of achieving faster training times by looking for the optimal set of algorithms for the chosen configuration of the model and inputs.

We investigated the complexities of the clinical translation of the proposed model. We utilized the large-scale fastMRI dataset for training and validation of the models whereas, in a clinical setting, the data distributions may be quite different. Although in this work we have attempted to illustrate the clinical compatibility of our model by presenting a series of experiments which introduce adversarial conditions such as higher accelerations, measurements noise, and anatomical abnormalities, more extensive experimentation and validation in a clinical setting are necessary. In our future work, we would be attempting to overcome these limitations while focusing more on replacing the fully supervised training setting with an unsupervised learning strategy in order to ease the dependency of the reconstruction model on prior data. That way, we would like to make our solutions more robust to out-of-distribution data which is a common concern in medical imaging. Also, we aim to extend our model to be compatible with multi-channel 3D MR data which could be beneficial in terms of exploiting global correlations across the slice dimension and coil sensitivities of the imaging system.

In conclusion, we have proposed a consolidated transformer-based solution, McSTRA, for accelerated MRI reconstruction which consists of several building blocks, i.e., the multi-branch feature extractor, cascaded network with intermediate loss computations, the magnitude image reconstruction tail, and the sampling-guided self-attention mechanism. Furthermore, it is important to note that our framework is a stand-alone transformer-based solution for MRI reconstruction, unlike the recent transformer-based works that incorporate some convolution layers within their model architectures in order to benefit from local feature aggregation. Both qualitative and quantitative analyses demonstrated the superior performance of McSTRA compared to state-of-the-art DL reconstruction methods. The results validate the potential of global feature extraction and reinforce the establishment of transformer-based DL models for high-quality MRI reconstruction.

## CRediT Authorship Contribution Statement

Mevan Ekanayake     : Conceptualization, Methodology, Software, Validation, Formal analysis, Investigation, Data Curation, Writing - Original Draft, Visualization
Kamlesh Pawar       : Resources, Writing - Review & Editing, Supervision
Mehrtash Harandi    : Resources, Writing - Review & Editing, Supervision
Gary Egan           : Writing - Review & Editing, Supervision, Funding acquisition
Zhaolin Chen        : Conceptualization, Resources, Writing - Review & Editing, Supervision, Funding acquisition

# Declaration of Competing Interest

None of the authors has any conflicts of interest to report

# Acknowledgments

This work was conducted as a part of the projects titled "Simultaneous to synergistic MR-PET: integrative brain imaging technologies" funded by the Australian Research Council Linkage Program (LP170100494) and "Biophysics-informed deep learning framework for magnetic resonance imaging" funded by the Australian Research Council Discovery Program (DP210101863).